# CUI-MET: Clinical Utility Index Dose Optimization Approach for Multiple-Dose, Multiple-Outcome Randomized Trial Designs

Fanni Zhang[a]*, Kristine Broglio[a], Michael Sweeting[b], Gina D'Angelo[a]

[a]Oncology Biometrics, AstraZeneca, Gaithersburg, USA

[b]GSK, London, UK

*Corresponding author: Fanni Zhang, fanni.zhang@astrazeneca.com, Oncology Biometrics, AstraZeneca, 1 MedImmune Way, Gaithersburg, Maryland, 20878, USA

Abstract: Dose optimization in oncology clinical trials has shifted from seeking the maximum tolerated dose to identifying the Optimal Biological Dose (OBD) that balances therapeutic benefits and risks across multiple clinical attributes. Existing advanced dose-finding methods can integrate multiple endpoints and compare dose levels but are often complex or computationally intensive, limiting their use in early-phase trials. To address these challenges, we propose the Clinical Utility Index Dose Optimization Approach for Multiple-dose Multiple-Outcome Randomized Trial Designs (CUI-MET). This framework integrates multiple binary endpoints using a clinical utility-based approach, calculating a combined clinical utility index (CUI) for each dose level by weighting endpoint responses. Both empirical and modeling methods can estimate marginal probabilities for each endpoint. These estimated probabilities are then combined using endpoint-specific weights to compute a utility score for each dose, and the dose with the highest score is selected as optimal. To enhance usability, we implemented these methods in an interactive R Shiny application and demonstrated their functionality through case examples. The framework's flexibility allows for different model selections and endpoint weighting schemes to reflect specific clinical priorities. Bootstrap analysis provides confidence intervals for the CUI and estimates the probability that each dose is selected as optimal, thereby evaluating the robustness of dose selection. By integrating multiple endpoints into a single utility index and incorporating user-friendly visualizations, CUI-MET offers a flexible and accessible solution for dose optimization in early-phase oncology trials, supporting informed decision-making and advancing patient-centered care.

Keywords: clinical utility index, dose optimization, multiple endpoints, R Shiny app

# 1. INTRODUCTION

In recent years, oncology drug development has evolved significantly, driven by the need for more effective and patient-centered therapeutic options. This includes a shift in what is required from the early phase studies designed to establish the recommended dose of new agents. Historically, the primary objective of phase I dose escalation trials was to determine the maximum tolerated dose (MTD) and phase II trials further evaluated that MTD for efficacy. This approach prioritizes maximizing efficacy with a ceiling on safety over achieving an optimal balance of both efficacy and safety. In practice this has resulted in the need for frequent dose reductions and delays for many agents due to safety and tolerability issues. There is now growing emphasis on identifying the Optimal Biological Dose (OBD) which maximizes efficacy while minimizing adverse effects through an explicit benefit-risk tradeoff. Regulatory guidance has reinforced this shift, encouraging randomized early phase trials with multiple dose levels (FDA 2024). These dose expansion trials highlight the necessity for refined decision-making strategies to compare limited dosing options effectively in the setting of small patient sample sizes that are typical of early phase oncology trials.

The Bayesian Optimal Interval (BOIN) design (Yuan, Hess et al. 2016) has been extended to consider both efficacy and safety (Ananthakrishnan, Lin et al. 2022), including BOIN12 (Lin, Zhou et al. 2020), BOIN-ET (Takeda, Taguri and Morita 2018) and U-BOIN (Zhou, Lee and Yuan 2019). The Dose-ranging Approach to Optimizing Dose (DROID) has been proposed to evaluate three endpoints: toxicity, efficacy, and a PD biomarker that is considered a surrogate for efficacy (Guo and Yuan 2023). While innovative, it introduces additional complexity through a two-stage structure. Even three endpoints may be unsatisfactory. A dose recommendation following an early phase trial can be multi-factorial, considering pharmacokinetics (PK), pharmacodynamics (PD), safety events of special interest, cumulative toxicity, and tolerability (D'Angelo, Gong et al. 2024).

Joint modeling has been explored in response to this need to integrate and directly compare mulitple different endpoints across randomized dose arms. Techniques such as multivariate regression models (Lin, Ryan et al. 2000) and structural equation models (Kessels, Moerbeek et al. 2021) simultaneously analyze multiple dependent variables and predictors, taking into account the correlations among the endpoints and provide a unified framework for estimation (Gelman 2021, Lin, Shi et al. 2022). While joint modeling provides a robust analytical foundation, extending these approaches to incorporate direct dose comparison and selection criteria is still necessary to fully address the question of optimal dosing in early-phase trials. Additionally, joint modeling often involves computational difficulties and can present challenges in interpreting the interdependencies among the endpoints.

Composite score approaches offer a promising solution in dose optimization (Song, Lin et al. 2013). This method simplifies analysis and interpretation by combining multiple different individual endpoints into a single score (Shaw 2018). These scores are typically constructed by assigning weights to each individual endpoint based on their relative importance to the overall clinical experience, and then combining them through various mathematical techniques. For instance, a simple additive or multiplicative model might be used, depending on the nature and scale of the endpoints involved.

Building upon the concept of composite scores, the clinical utility index (CUI) offers a more refined approach to integrating multiple clinical endpoints. Originally designed to assess the therapeutic index of new drugs, the CUI is defined as an integrated measure of risk-benefit across multiple individual endpoints that can be quantified for the entire exposure range (Rowland M. 1980, Khan, Perlstein and Krishna 2009, Ouellet, Werth et al. 2009). Other utility-based approaches such as BOIN12 (Lin, Zhou et al. 2020) and U-MET (D'Angelo, Gong et al. 2024), combine endpoints jointly. Specification of the joint distribution of endpoints becomes very difficult as the number of endpoints increases beyond 2 or 3. The CUI method models the marginal distribution of each endpoint since it is typically difficult to reliably estimate a multivariate correlation structure from small sample sizes and similar operating characteristics have previously been observed for independent compared to jointly modelled outcomes in Phase I/II trials (Cunanan and Koopmeiners 2014). This allows for straightforward calculations of the score using an arithmetic weighted average (Song, Lin et al. 2013, Winzenborg, Soliman and Shebley 2021) or a geometric mean of the independent expected outcomes, which are often rescaled to reflect desirability scores scaled between 0 and 1 (Coffey, Gennings and Moser 2007), thus easily accommodating more than 2 or 3 individual endpoints.

We introduce the clinical utility index dose optimization approach for multiple-dose randomized trial designs (CUI-MET). The CUI-MET addresses both the complexities of endpoint integration and the critical need for comparative dose analysis. This innovative framework integrates a flexible number of endpoints and facilitates statistical comparisons between dose levels. The approach is flexible in allowing both independent (or empirical) modelling of outcome responses across the dose-levels and the incorporation of parametric modelling, using a suite of four commonly used dose-response models (logit linear, logit quadratic, Emax and exponential models). By using a utility-based approach, the CUI-MET allows a customized balance of efficacy, safety, tolerability, and PK/PD metrics, supporting a more informed, nuanced, and real-world decision-making process in determining the OBD. To increase accessibility and user-friendliness, we have developed an intuitive R Shiny application. This tool offers clinicians and researchers a clear graphical presentation of dose-endpoint relationships alongside composite CUI scores.

The remainder of this paper is organized as follows. In Section 2, we propose the CUI-MET detailed with both empirical and modeling approaches, and introduce the corresponding R Shiny application. In Section 3, we illustrate our methods with examples and present results. We conclude with a brief discussion in Section 4.

## 2. METHODS

We assume an early phase clinical trial that will randomize $N$ patients across a set of ordered dose levels $j = 1, 2, …, J$ with $K$ individual binary endpoints of interest for dose optimization. Let $Y_{ijk}$ represent the binary outcome value of the $k^{th}$ endpoint $EP_k$ for the $i^{th}$ patient assigned to dose level $j$. We assume for consistency across all $K$ endpoints such that $EP_k = 1$ indicates a positive outcome. Thus a larger CUI value indicates a better overall result for that dose. Our primary aim is to identify the dose with the largest CUI.

We adopt a weighting scheme similar to that used by Winzenborg et al (Winzenborg, Soliman and Shebley 2021). Each endpoint is assigned its own weight, $w_1, w_2, …, w_K$, within the range of 0 to 5 with higher values indicating greater importance for a particular endpoint. These weightings are then normalized to $\widetilde{w_1}, \widetilde{w_2}, …, \widetilde{w_K}$ so that their sum equals one, ensuring balanced contributions to the overall utility calculation. The normalization is calculated as follows:

$$\widetilde{w_k} = \frac{w_k}{\sum_{l=1}^{K} w_l}, \ k = 1, 2, …, K.$$

Using these normalized weights, we define the combined CUI for each dose level $j$ as the utility weighted mean (UWM):

$$CUI(j) = UWM_j = \sum_{k=1}^{K} \widetilde{w_k} P(EP_k = 1 | dose\ j)$$

If endpoint weights are not applied, the CUI can be simplified to the utility mean (UM), computed as the average probability of a positive outcome across all endpoints:

$$CUI(j) = UM_j = \frac{1}{K} \sum_{k=1}^{K} P(EP_k = 1 | dose\ j)$$

The UWM reduces to the UM when all endpoints are weighted equally, reflecting a scenario where each endpoint is considered of equal importance in the analysis. The optimal dose is then identified as the dose level with the highest UWM when using the UWM metric or the highest UM when using the UM metric.

## 2.1 Empirical Method

In the empirical method, the marginal probability of a positive outcome for endpoint $EP_k$ at a specific dose level $j$ is estimated using the observed mean response rate. This is calculated by taking the average of the binary outcomes for all patients assigned to that dose level. The probability $P(EP_k = 1 | dose\ j)$ is thus given by:

$$P(EP_k = 1 | dose\ j) = \frac{1}{N} \sum_{i=1}^{N} Y_{ijk}$$

This method straightforwardly aggregates patient outcomes to yield a simple and clear measure of the observed clinical utility of each dose level. However, the method does not borrow information about the outcomes from other dose levels. This may be important if there is a known dose-response relationship, for example a monotonically non-decreasing probability of the outcome occurring as dosage increases.

## 2.2 Modeling Method

We additionally include four choices for parametric modeling across dose levels to estimate the marginal probability for each endpoint: linear and quadratic regression models on the logit scale, the Emax model, and the exponential model (Dette, Kiss et al. 2010, Pinheiro, Bornkamp et al. 2014). Non-monotonic associations are permitted when using a logit linear or logit quadratic model. By definition, the Emax and exponential models inherently assume monotonic relationships due to their mathematical structure. Each endpoint is modeled independently and the model-estimated marginal rates are then used in the CUI calculation. For simplicity and clarity in the modeling descriptions below, we use $P(EP = 1)$ instead of $P(EP_k = 1)$ to represent the marginal probability of a positive outcome for a generic endpoint.

For the logit linear and logit quadratic models, predicted probabilities are estimated by fitting a logistic regression model. Monotonicity constraints can be optionally applied. In the linear case, the slope parameter is exponentiated to ensure positivity. In the quadratic model, constraints are applied to the derivative of the fitted curve across the dose range, penalizing any violations of monotonicity to ensure a non-decreasing response. These models were implemented using custom likelihood functions and optimization routines in R.

For the Emax and exponential models, we use a two-stage modeling approach following the generalized MCP-Mod methodology implemented in the *DoseFinding* R package (Bornkamp, Pinheiro et al. 2025). In the first stage, a generalized linear model (GLM) without an intercept is fit to estimate dose-specific log-odds of a positive response, along with their covariance matrix. In the second stage, these estimates are used to fit a parametric dose-response curve (Emax or exponential) on the logit scale, and the predicted values are then transformed to the probability scale for use in the CUI calculation. This approach is consistent with the recommended procedure for binary data modeling, as outlined in the package vignette and associated reference (Pinheiro, Bornkamp et al. 2014).

These modeling approaches were selected for their ablity to fit dose-response relationships across various clinical endpoints. The choice of model depends on the underlying biological assumptions about how the treatment dosage affects the endpoint and the observed data patterns.

### 2.2.1 Logit linear model

The logit linear model is expressed as:

$$\text{logit}\big(P(EP = 1)\big) = \beta_0 + \beta_1 * dose$$

This model describes a simple linear relationship between the log odds of a positive outcome for an endpoint by dose.

### 2.2.2 Logit quadratic model

The logit quadratic model adds a quadratic term to capture curvature in the dose-response relationship:

$$\text{logit}\big(P(EP = 1)\big) = \beta_0 + \beta_1 * dose + \beta_2 * dose^2$$

This model is useful for exploring more complex dose-response shapes that cannot be adequately described by a simple linear model.

### 2.2.3 Emax model

The Emax model is formulated as:

$$\text{logit}(\hat{p}) = E_0 + E_{max} * \frac{dose}{ED_{50} + dose}$$

In this model, $\hat{p}$ is the predicted probability of a positive response from the first-stage GLM fitting, $E_0$ is the baseline effect, $E_{max}$ represents the maximum achievable effect above the baseline, and $ED_{50}$ is the dose at which half of the maximum effect is achieved. The Emax model is particularly valued for its ability to describe saturation effects as dose increases.

### 2.2.4 Exponential model

The exponential model is given by:

$$\text{logit}(\hat{p}) = E_0 + E_1(\exp(dose/\sigma) - 1)$$

Here, $E_1$ controls the magnitude of the dose effect, $\sigma$ is a scale parameter that influences both the steepness and convexity of the dose-response curve.

## 2.3 Bootstrap Confidence Intervals

Formal hypothesis testing is not required nor feasibile in the dose optimization setting with very small sample sizes. However, a critical step in identifying the OBD is to understand the uncertainty associated with CUI estimates when comparing doses. We accomplish this through a bootstrap resampling method. The bootstrap is a non-parametric statistical technique that involves repeatedly sampling with replacement from the observed data to create numerous pseudo-datasets, known as bootstrap samples. This approach allows for the estimation of the sampling distribution of a statistic without relying on strict parametric assumptions, making it particularly suitable for complex models or situations where the theoretical distribution is unknown.

In our study, we generate 1,000 bootstrap samples by resampling patients with replacement within each dose level group (i.e. stratified bootstrapping), maintaining the same sample size per dose level as in the original dataset. This was implemented using the R package *boot* (Canty and Ripley 2024). For each bootstrap sample, we re-estimated the marginal probabilities for each endpoint and dose level using the specified method and calculate the UWM and UM as CUI scores. With a significance level of α=0.05, the two-sided $(1-\alpha)*100\%$ confidence intervals of the UWM and UM for each dose level are calculated using the percentile bootstrap method. Specifically, the lower and upper bounds of the confidence intervals were determined as the $100*(\alpha/2)$ and $100*(1-\alpha/2)$ percentiles of the bootstrap estimates, respectively.

Additionally, we assess the robustness of dose selection by calculating the percentage of times each dose was identified as the highest UWM or UM dose across the 1,000 bootstrap samples. This provided insight into the possible uncertainty surrounding the optimal dose recommendation under the variability captured by the bootstrap resampling.

## 2.4 R Shiny Application

We developed an interactive R Shiny application for the implementation of the methods described above in 2.1 and 2.2. This user-friendly tool serves as an interactive and dynamic platform for dose optimization.

Key features of the CUI-MET Shiny app include:

- Interactive Interface: Features a user-friendly interface with dynamic visuals and tables that facilitate the exploration and analysis of clinical trial data.
- Integration of Multiple Endpoints: Enables the input and analysis of data from a flexible number of endpoints.
- Calculation of Clinical Utility Metrics: Calculates both the utility mean (UM) and the utility weighted mean (UWM) for the CUI, providing different views of the tradeoffs between each dose's potential benefits and risks.
- Optimal Dose Identification: Identifies the dose with the highest CUI by the selected methods and designates that dose as the OBD.

- Bootstrap Analysis: Generates bootstrap confidence intervals around the CUI and determines the probability of each dose being the OBD, offering a measure of the robustness of the dose-optimization decisions.

- Dose Comparison and Ranking: Allows users to compare and rank different doses based on the CUI scores, aiding in the visual and statistical assessment of dose efficacy and safety.

Figure 1 provides a screenshot of the Shiny application interface, illustrating the input panel (left) where users specify endpoints, model types, and weights, and the output panel (right), which displays calculated utility scores, selected doses, and related visualizations.

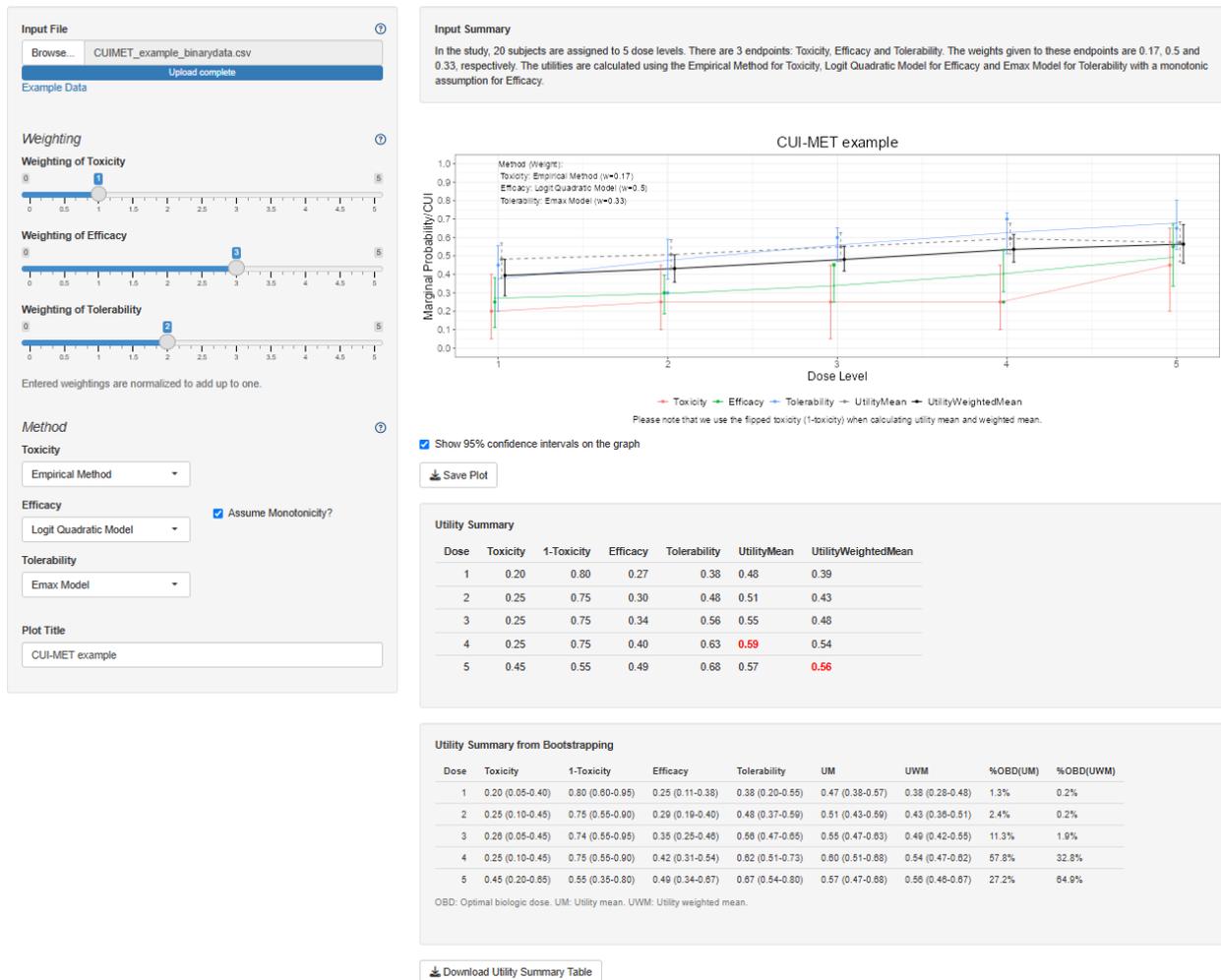

*Figure 1. Screenshot of the CUI-MET application interface.*

**Input Requirements**

The input data should be formatted as a .csv file. Users can refer to an example dataset accessible via the "Example Data" hyperlink to understand the required data structure. Essential input variables for the functionality of the shiny app include:

- ID: Patient ID number
- Dose: A numeric variable with values ranging from 1 to the highest dose level used in the study. These values represent ordered dose levels and are treated as categorical levels within the application. Users should provide integer values (e.g., 1, 2, ..., $J$) to indicate distinct dose levels, where the order reflects increasing dosing.
- Toxicity: Binary values (0 or 1), with 1 representing patients who experienced toxicity, and 0 representing patients who did not experience toxicity during the study.
- Efficacy: Binary values (0 or 1), where 1 denotes patients with positive efficacy outcomes, while 0 denotes patients without positive efficacy outcomes during the study.

Additional binary outcomes (0 or 1) can also be imported and the corresponding user-specified variable names can be automatically extracted from user's input data and displayed in the weighting parameter labels. Because of the assumed convention that an outcome of 1 indicates a positive outcome, the app automatically converts the toxicity endpoint to 1– Toxicity, modeling the probability of not experiencing an event. However, for any other similar variables, such as tolerability where 1 might typically indicate experiencing a tolerability event, users should make this transformation in the input data in order to model the probability of a positive event, i.e. not having a tolerability event.

**Input Parameter Configuration**

Upon uploading the data, weighting sliders for each endpoint appear, ranging from 0 to 5 with a default setting of 1, adjustable in 0.1 increments. Each endpoint has its own weighting slider, providing custom, flexible, and fine-grained weighting adjustments for each of the multiple endpoints. Dynamic adjustment of the weighting sliders within the app is allowed. Users should tailor the weighting of each endpoint according to their clinical setting. The methods for modeling each endpoint are available via dropdown menus, with option to specify monotonicity assumptions for endpoints analyzed using logit models. This option is available for both the logit linear and logit quadratic models, and applies to all endpoints except toxicity, for which a monotonic increase in probability with dose is always assumed. When enabled, the monotonicity constraint ensures that estimated response probabilities are non-decreasing across increasing dose levels.

**Output Overview**

The output panel displays a comprehensive summary of the input data, analytical results, and corresponding visualizations tailored according to the user-defined parameters. It functions as a reactive output interface, automatically updating to reflect changes whenever adjustments are made to the input weightings or method selections. This feature ensures that users can immediately see the impact of their adjustments on the results and visualizations, facilitating an interactive and iterative approach to analyzing the trial data.

# 3. RESULTS

**Overview of Case Examples**

In this section, we present three simulated example datasets to demonstrate the application of the CUI-MET. Each example dataset includes 5 dose levels with 30 patients treated at each dose and 3 key binary endpoints: toxicity, efficacy, and tolerability.

The data was simulated using a multivariate normal distribution, with predefined dose-specific probabilities assigned to each endpoint. Continuous samples are generated and subsequently transformed into binary outcomes (0/1) by comparing them to the inverse normal cumulative distribution function corresponding to the specified probabilities. We consider a covariance structure that allows the endpoints to be correlated through the specification of off-diagonal elements in the covariance matrix. For the main examples presented in this paper, we assume independence between endpoints to simplify interpretation. We also explored correlated scenarios to assess robustness and found that the results were generally consistent. This simulation framework allows for the generation of realistic relationships between the endpoints of interest, whether independent or correlated.

Each example dataset demonstrates distinct observed dose-response patterns across the endpoints. In Example 1, all endpoints increase with dose, while Example 2 shows plateau effects for efficacy and tolerability at higher doses. Example 3 presents a more complex relationship, where efficacy follows a concave pattern, initially increasing but later decreasing, while tolerability declines approximately linearly. Figure 2 shows the empirical marginal probabilities of each endpoint across dose levels for each example.

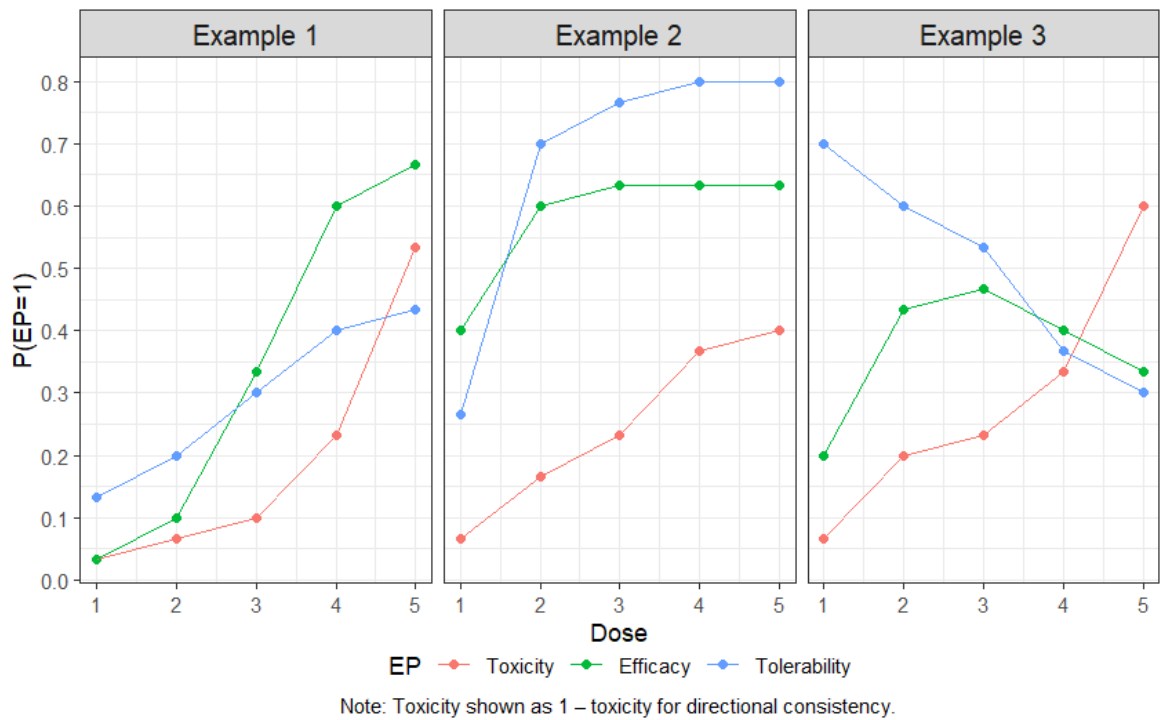

Note: Toxicity shown as 1 – toxicity for directional consistency.

*Figure 2. Empirical marginal probabilities of each endpoint (K=3) across dose levels (J=5) based on observed data from N=30 subjects for each dose level in each example dataset. For visualization purposes, toxicity is shown as 1 – toxicity (i.e., the probability of no toxicity), so that higher values indicate more desirable outcomes and align directionally with efficacy and tolerability.*

**Modeling Approach Selection and Weighting Schemes**

Within each example, we apply the various parametric modeling approaches to estimate the probabilities of a positive outcome for each endpoint. The selection of models was guided by visual inspection of the dose-response trends in each scenario (Figure 2), supported by biological plausibility and simplicity of fit. In Example 1: 1) an exponential model was chosen for toxicity due to its non-linear increase, particularly at higher dose levels, 2) a logit quadratic model was deemed appropriate for efficacy because of its slightly non-linear trend, and 3) a logit linear model was considered for tolerability given the roughly linear increase. In Example 2: 1) an Emax model was selected for efficacy and tolerability to capture the plateau effects observed at higher doses, and 2) a logit-linear model was used for toxicity, given its approximately linear association across dose levels. Lastly, in Example 3: 1) a logit quadratic model was applied to fit the concave pattern of efficacy, 2) an exponential model was used for toxicity to account for the sharp non-linear increase at higher doses, and 3) a logit linear model was chosen for tolerability due to its steady decline. These modeling choices are shown in Table 1. These model selections are relatively subjective and primarily serve to demonstrate the tool's functionality. In practice, users can refine model selection through an a priori understanding of the clinical plausability and by comparing the fit of different models using graphical assessments, depending on the specific needs and preferences of their study.

Two different weighting schemes are considered within each example dataset. The weighting schemes were selected to reflect different clinical priorities and are shown in Table 1. In general, the first weighting scheme gives efficacy the highest weight whereas the second weighting scheme lessens the weight on efficacy and increases the weight given to safety and/or tolerability.

Example 1: In weight scheme 1, efficacy was given the highest priority (weight 0.5), while toxicity (weight 0.2) was a concern as the dose increased. Tolerability (weight 0.3) was moderately important to account for patient comfort. In weight scheme 2, greater emphasis was placed on tolerability (weight 0.5), reflecting a scenario where patient experience was critical. Efficacy (weight 0.2) and toxicity (weight 0.3) were still considered, but with less weight.

Example 2: Weighting scheme 1 is the same as described above. In weight scheme 2, equal emphasis is placed on both toxicity and efficacy, each with a weight of 0.4. This reflects a scenario where efficacy and toxicity remain important, but minimizing tolerability is given equal priority to ensure patient safety.

Example 3: Weighting scheme 1 is the same as described above. In weight scheme 2, toxicity (weight 0.4) and tolerability (weight 0.4) were equally weighted. Efficacy (weight 0.2) was given less weight, with a greater focus on managing safety and side effects.

*Table 1. Modeling approach selection for each endpoint across examples and corresponding weight schemes*

| Example | Endpoint | Model | Weight Scheme 1 | Weight Scheme 2 |
|---------|----------|-------|-----------------|-----------------|
| Example 1 | Toxicity | Exponential | 0.2 | 0.3 |
| | Efficacy | Logit Quadratic | 0.5 | 0.2 |
| | Tolerability | Logit Linear | 0.3 | 0.5 |
| Example 2 | Toxicity | Logit Linear | 0.2 | 0.4 |
| | Efficacy | Emax | 0.5 | 0.4 |
| | Tolerability | Emax | 0.3 | 0.2 |
| Example 3 | Toxicity | Exponential | 0.2 | 0.4 |
| | Efficacy | Logit Quadratic | 0.5 | 0.2 |
| | Tolerability | Logit Linear | 0.3 | 0.4 |

**Results for CUI and OBD**

To compare different weighting schemes, we presented the resulting CUI-MET from our R Shiny app in Figure 3. These same plots, but incorporating the bootstrap CIs, are provided in the appendix (Figure A1). Each plot displays empirical results for each endpoint across the five dose levels as dots and the modeled estimates with lines. The UM is shown in each example, and because the UM always considers each endpoint with equal weight, it remains unchanged across the weighting schemes considered. In contrast, the UWM line does incorporate the endpoint-specific weights and by comparing the UWM across figures, we can examine the sensitivity of the results to the different weighting choices. Across all examples, dose levels were ranked and compared based on these CUI values, with the OBD identified as the dose level with the highest CUI.

In Example 1, efficacy is increasing across all doses and so with weight scheme 1, which gives efficacy the highest weight, dose level 5 is selected as the OBD. However, dose level 5 also sees much higher toxicity and so under weighting scheme 2, dose level 4 was selected as the OBD. Equally weighting all endpoints with the UM, dose level 4 was identified as the OBD (Figure 3 and Table 2). Bootstrap estimates for UWM further supported these findings, with dose level 5 selected as optimal 62.6% of the time in weight scheme 1, while dose level 4 was identified as optimal with a frequency of 71% in weight scheme 2 (Table 3).

In Example 2, efficacy plateaus starting with dose level 3 and yet toxicity and tolerability events continue to increase. Weighting scheme 1 mirrors the clinical reality that increasing the doses no longer brings additional benefit but does bring additional risk. Weighting scheme 1 estimated similar CUI for dose levels 3 and 4 and a lower CUI for dose level 5. In weight scheme 2, where toxicity and efficacy were given equal emphasis, there are larger decreases in the UWM across doses 4 and 5. These larger declines reflect the weighting scheme's concern with rising toxicity quickly outweighing the efficacy. Based on UWM, dose level 4 in weight scheme 1 and dose level 3 in weight scheme 2 each emerged as the top-ranked dose within their respective weighting schemes, with only a slight advantage over nearby doses (Table 2). Bootstrap results show that weighting scheme 1 selects dose level 4 with 55% probability, but weighting scheme 2 selects dose level 3 with 73.9% probability highlighting more certainty for the OBD under weighting scheme 2 (Table 3).

In Example 3, efficacy takes a concave shape and so for weighting scheme 1, the UWM peaks at dose level 2, and then begins to decline across doses 3 through 5 (Figure 3 and Table 2). Dose levels 2 and 3 are estimated as having very similar values, reflected by the bootstrap sampling selecting dose 2 44.5% of the time and dose 3 46% of the time (Table 3), indicating only a marginal difference in selection frequency. In weight scheme 2, where equal emphasis was placed on toxicity and tolerability, the UWM was highest at dose level 1 reflecting the increasing toxicity and tolerability across the doses, and declined more sharply at higher doses where efficacy also declines. This pattern reflected the increased emphasis on toxicity and tolerability, leading to a preference for the lowest dose level to minimize adverse effects. Under weight scheme 2, dose level 1 was the preferred choice, being selected 85.8% of the time (Table 3).

Overall, the bootstrapped results reflected the dose ranking determined by UWM but quantified the confidence, or lack of, in that choice. The top-ranked dose based on UWM closely aligned with the most frequently selected dose in the bootstrap analysis. In scenarios where multiple doses had similar UWM scores, bootstrap resampling provided additional insight by highlighting uncertainty and revealing the most consistently selected dose across resamples. The stability of the optimal dose selection across resampling suggests that the CUI-MET framework reliably captures the trade-offs between efficacy, toxicity, and tolerability in dose optimization. These results highlight the robustness of the approach in adapting to different clinical priorities while maintaining a consistent ranking of preferred doses.

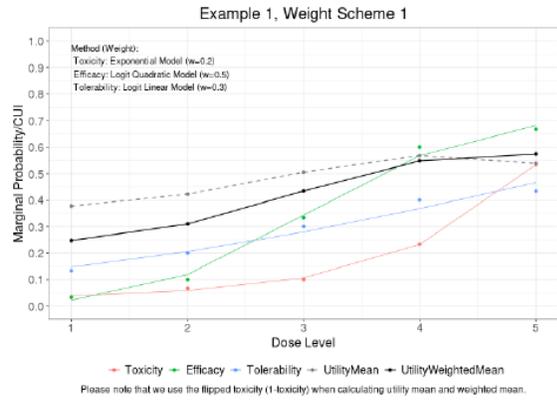
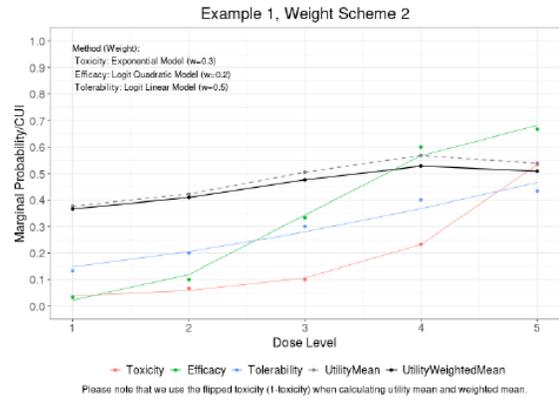
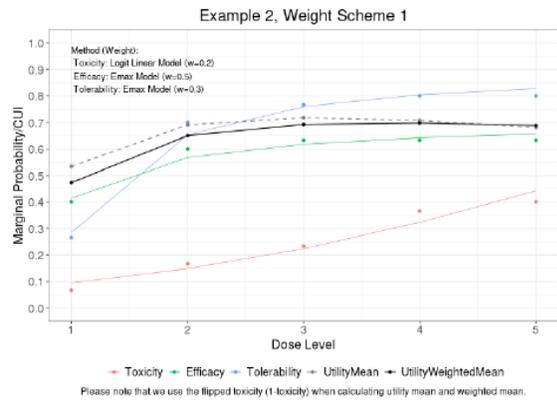
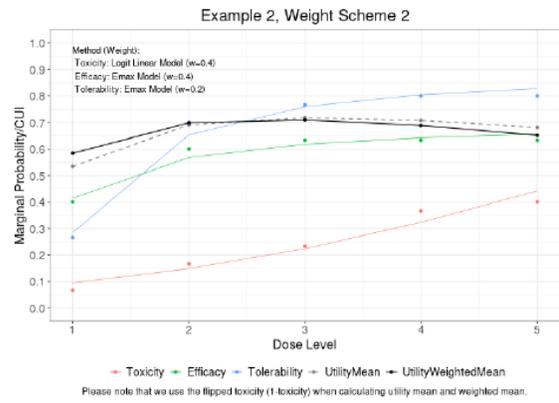
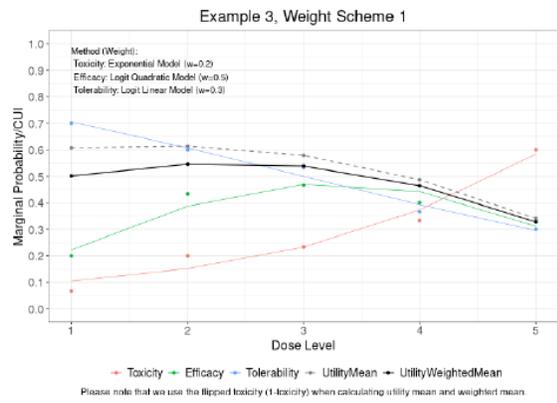
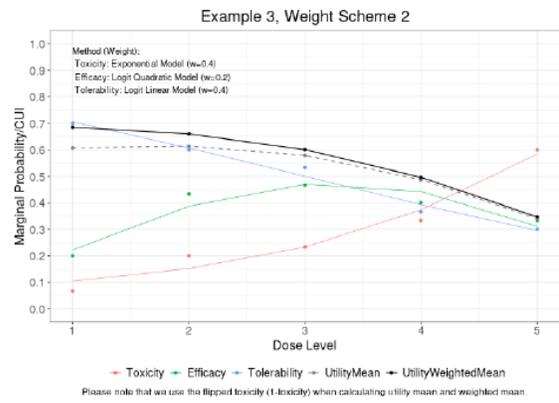

*Figure 3. CUI-MET plots for each weight scheme across examples*

*Table 2. Clinical utility index summary*

| Example | Dose | Toxicity P(Tox=1) | 1-Toxicity 1-P(Tox=1) | Efficacy P(Eff=1) | Tolerability P(Tol=1) | UM | Weight Scheme 1 UWM | Weight Scheme 2 UWM |
|---|---|---|---|---|---|---|---|---|
| Example 1 | 1 | 0.038 | 0.962 | 0.022 | 0.147 | 0.377 | 0.248 | 0.367 |
| | 2 | 0.058 | 0.942 | 0.119 | 0.206 | 0.422 | 0.310 | 0.409 |
| | 3 | 0.106 | 0.894 | 0.343 | 0.280 | 0.506 | 0.434 | 0.477 |
| | 4 | 0.232 | 0.768 | 0.568 | 0.368 | 0.568 | 0.548 | 0.528 |
| | 5 | 0.533 | 0.467 | 0.681 | 0.466 | 0.538 | 0.574 | 0.509 |
| Example 2 | 1 | 0.095 | 0.905 | 0.414 | 0.285 | 0.534 | 0.473 | 0.584 |
| | 2 | 0.149 | 0.851 | 0.568 | 0.653 | 0.691 | 0.650 | 0.698 |
| | 3 | 0.224 | 0.776 | 0.618 | 0.760 | 0.718 | 0.692 | 0.710 |
| | 4 | 0.323 | 0.677 | 0.642 | 0.804 | 0.708 | 0.698 | 0.688 |
| | 5 | 0.442 | 0.558 | 0.657 | 0.828 | 0.681 | 0.688 | 0.652 |
| Example 3 | 1 | 0.105 | 0.895 | 0.223 | 0.705 | 0.607 | 0.502 | 0.684 |
| | 2 | 0.152 | 0.848 | 0.387 | 0.607 | 0.614 | 0.545 | 0.660 |
| | 3 | 0.233 | 0.767 | 0.470 | 0.500 | 0.579 | 0.538 | 0.601 |
| | 4 | 0.374 | 0.626 | 0.442 | 0.393 | 0.487 | 0.464 | 0.496 |
| | 5 | 0.583 | 0.417 | 0.312 | 0.295 | 0.341 | 0.328 | 0.347 |

*Note: P(Tox=1), P(Eff=1), and P(Tol=1) represent the estimated mariginal probabilities: P(Toxicity=1), P(Efficacy=1), and P(Tolerability=1). UM=Utility Mean. UWM=Utility Weighted Mean.*

*Table 3. Clinical utility index summary from bootstrapping*

| Example | Dose | UM (95% Bootstrap CI) | %OBD (UM) | Weight Scheme 1 | | Weight Scheme 2 | |
|---|---|---|---|---|---|---|---|
| | | | | UWM | %OBD (UWM) | UWM | %OBD (UWM) |
| Example 1 | 1 | 0.37 (0.33-0.41) | 0% | 0.24 (0.21-0.28) | 0% | 0.36 (0.32-0.41) | 0% |
| | 2 | 0.41 (0.37-0.46) | 0% | 0.30 (0.25-0.35) | 0% | 0.40 (0.36-0.45) | 0% |
| | 3 | 0.50 (0.45-0.56) | 0% | 0.43 (0.37-0.50) | 0% | 0.47 (0.43-0.52) | 0.30% |
| | 4 | 0.57 (0.51-0.64) | 77.90% | 0.55 (0.48-0.62) | 37.40% | 0.53 (0.47-0.60) | 71% |
| | 5 | 0.54 (0.44-0.63) | 22.10% | 0.57 (0.46-0.68) | 62.60% | 0.51 (0.41-0.60) | 28.70% |
| Example 2 | 1 | 0.53 (0.45-0.62) | 0% | 0.47 (0.38-0.58) | 0% | 0.58 (0.50-0.67) | 0% |
| | 2 | 0.68 (0.63-0.73) | 0.80% | 0.64 (0.59-0.69) | 0.10% | 0.69 (0.64-0.74) | 16% |
| | 3 | 0.71 (0.67-0.76) | 71.20% | 0.68 (0.64-0.74) | 19% | 0.70 (0.66-0.75) | 73.90% |
| | 4 | 0.70 (0.65-0.75) | 24.60% | 0.69 (0.64-0.75) | 55.10% | 0.69 (0.63-0.74) | 8.90% |
| | 5 | 0.68 (0.62-0.75) | 3.40% | 0.69 (0.62-0.76) | 25.80% | 0.65 (0.58-0.73) | 1.20% |
| Example 3 | 1 | 0.60 (0.52-0.67) | 32.40% | 0.50 (0.42-0.59) | 9.40% | 0.67 (0.60-0.74) | 85.80% |
| | 2 | 0.61 (0.55-0.67) | 59.30% | 0.54 (0.47-0.61) | 44.50% | 0.65 (0.59-0.71) | 13.90% |
| | 3 | 0.58 (0.52-0.64) | 8.30% | 0.54 (0.47-0.60) | 46% | 0.60 (0.54-0.66) | 0.30% |
| | 4 | 0.50 (0.43-0.56) | 0% | 0.47 (0.40-0.53) | 0.10% | 0.51 (0.44-0.57) | 0% |
| | 5 | 0.33 (0.25-0.41) | 0% | 0.32 (0.24-0.41) | 0% | 0.34 (0.26-0.42) | 0% |

*Note: UM (95% Bootstrap CI) is the Utility Mean along with its 95% percentile confidence interval (CI) from bootstrapping. %OBD (UM) and %OBD (UWM) represent the percentage of times a dose was optimal based on Utility Mean and Utility Weighted Mean, respectively.*

# 4. DISCUSSION

We proposed the CUI-MET approach and developed a user-friendly R shiny application to enable the implementation of this method for dose optimization needs in oncology. By integrating multiple binary endpoints into a single utility index, CUI-MET facilitates comprehensive evaluation and comparison of the clinical tradeoffs associated with different dose levels. Our simulated examples demonstrated the flexibility and effectiveness of CUI-MET in identifying the OBD across different dose-response patterns and clinical priorities. The ability to customize the endpoints included as well as the weighting schemes allows for varying clinical settings and considerations. Parametric models to estimate endpoint probabilities further adapt the framework to different data patterns and underlying biological assumptions.

Our approach has several key advantages over the existing approaches. Compared to traditional dose-finding methods that may integrate multiple endpoints but often involve complex computations or require substantial computational resources, CUI-MET offers a more accessible and practical approach, particularly to the non-statistical audience. By combining multiple endpoints into a composite score using straightforward calculations, it aligns with the current regulatory emphasis that dosing should balance therapeutic benefits with potential risks. This is a "totality of the data" approach that is also explicit and transparent in how different aspects of the data are valued relative to each other. We include bootstrap resampling to quantify the uncertainty associated with utility estimates and OBD selection. This feature is particularly valuable in early-phase trials with small sample sizes. In addition to selecting the dose with the highest utility score, the bootstrap results also provide the probability that each dose is optimal. This allows users to incorporate a level of evidence into their decision-making, such as selecting a dose only if its probability of being optimal exceeds a predefined threshold. Finally, the development of the interactive R Shiny application further strengthens the practical use of CUI-MET. Its dynamic visualization capabilities offer clear and intuitive representations of relationships between endpoints, dose levels, and clinical utility scores. This tool aids in the interpretation and communication of results, making the framework accessible to clinicians and researchers without extensive statistical programming expertise.

Despite these strengths, several limitations warrant discussion. The app currently only includes binary endpoints. Extending the methodology to incorporate continuous or time-to-event endpoints could further enhance its utility. Survival endpoints are particularly important in oncology trials, as they are typically the clinical endpoints of interest. Incorporating these data types into the CUI-MET framework would allow for a more comprehensive assessment of treatment effects and could improve the relevance of OBD selection in clinical practice.

Additionally, CUI-MET framework accounts for marginal probabilities rather than joint combinations of endpoints. While this simplifies weight assignment and interpretation, a joint weighting approach may better capture the patient-level tradeoffs between benefits and risks. Another limitation lies in the lack of formal statistical methods for direct dose comparison. While CUI-MET provides an aggregated utility score for each dose level, future work could involve integrating formal hypothesis testing procedures to compare clinical utility scores, adjusting for multiple comparisons, or applying confidence intervals to assess the significance of differences between doses. Recent work on the UMET framework has introduced a Bayesian hypothesis testing approach to formally compare utility scores across doses using probability thresholds to identify the OBD (D'Angelo, Chen and Ran 2025). Future extensions of CUI-MET could leverage this testing approach to strengthen the robustness of dose selection by offering statistical evidence to support the OBD selection.

The examples presented are based on simulated data, but we have not conducted a formal simulation study to assess operating characteristics of CUI-MET, which is a topic for future research. Exploring the application of CUI-MET in actual clinical datasets would help illustrate its use and impact on dose selection in real-world settings. Opportunities for further improvement include integrating other clinically relevant endpoint types and adding trial design elements. Trial design elements could include assessing the average performance of CUI-MET under different assumed true scenarios as well as adaptive design elements such as adaptive randomization or arm dropping in order to treat patients in the trial on the best doses and gather additional information on the dose levels that emerge as the most likely candidates. These advancements could enhance trial efficiency and address ethical considerations.

In summary, CUI-MET represents a significant advancement in dose optimization methodologies for multiple-dose randomized trial designs. By integrating multiple endpoints into a unified utility index and providing systematic dose comparisons, it supports clinically relevant and patient-centered decision-making in oncology clinical trials. The incorporation of additional endpoint types and trial design features in future iterations will further strengthen its utility. The accompanying R Shiny application facilitates its adoption in clinical research, potentially improving therapeutic outcomes for patients.

Appendix

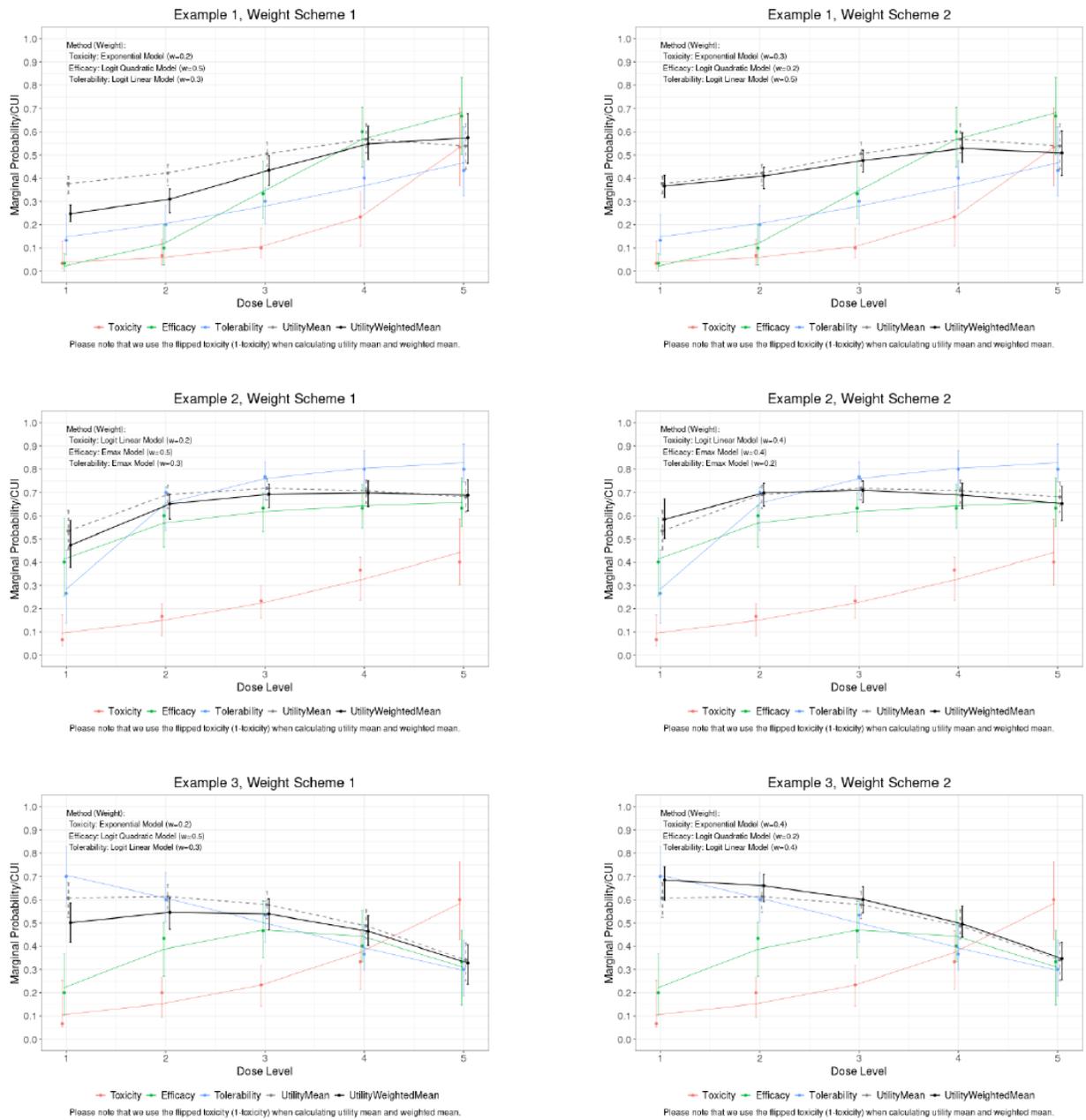

*Figure A1. CUI-MET plots with 95% bootstrap confidence intervals for each weight scheme across examples*